# A universal and improved mutation strategy for iterative wavefront shaping

Hui Liu[1], Xiangyu Zhu[1], Xiaoxue Zhang[1], Yongquan Liao[1], Xudong Chen[1]*, Zhili Lin[1]**

[1]Fujian Key Laboratory of Light Propagation and Transformation, College of Information Science and Engineering, Huaqiao University, Xiamen, Fujian 361021, China
*Corresponding author: chenxd@hqu.edu.cn; ** corresponding author: zllin2008@gmail.com

Recent advances in iterative wavefront shaping (WFS) techniques have made it possible to manipulate the light focusing and transport in scattering media. To improve the optimization performance, various optimization algorithms and improved strategies have been utilized. Here, a novel guided mutation (GM) strategy is proposed to improve optimization efficiency for iterative WFS. For both phase modulation and binary amplitude modulation, considerable improvements in optimization effect and rate have been obtained using multiple GM-enhanced algorithms. Due of its improvements and universality, GM is beneficial for applications ranging from controlling the transmission of light through disordered media to optical manipulation behind them.
**Keywords**: wavefront shaping, iterative optimization algorithm, scattering media, spatial light modulator.
DOI: COLXXXXXX.XXXXXX.

## 1. Introduction

When light propagates through complex media, such as biological tissues and multimode fibers, refractive index inhomogeneity causes multiple scattering and distortion. This phenomenon is usually seen as obstacles for biomedical imaging, telecommunications, and photodynamic therapy. As an effective method, iterative wavefront shaping (WFS) is capable of manipulating the incident wavefront and compensate the wavefront distortion due to multiple scattering [1]. Over the years, various iterative algorithms have been reported to achieve light focusing after scattering media, such as genetic algorithm (GA) [2, 3], particle swarm optimization (PSO) algorithm [4, 5], simulated annealing (SA) algorithm [6, 7] and ant colony (ACO) algorithm [8] . Furthermore, a set of hybrid [9, 10] or improved strategies [11-16] are proposed to enhance the optimization efficiency or adaptability. Among these strategies, a mutation procedure is widely suggested to expand diversity, including single point mutation [11, 12], decaying mutation [13], dynamic mutation [14-16]. These mutation operations concentrate on determining the mutation number to enhance optimization performance. However, the mutated pixel value still relies on random selection, which always lead to ineffective or redundant measurements. In addition, most improved strategies are only feasible for a specified algorithm. To date, an improved strategy that is suitable for various algorithms has not been demonstrated yet.

In this letter, we propose a universal and improved mutation strategy with guidance for iterative WFS. Instead of completely random selection employed in existing improvement strategies, a guided mutation (GM) operation is utilized to estimate the mutated pixel value and number. Moreover, not limited to a specific algorithm, we extend GM to various feedback-based WFS algorithms. With the enhancement of GM, commonly-used iterative algorithms for WFS, such as GA, PSO, ACO and SA, are demonstrated to obtain greater optimization performance both in simulation and experiment. In addition to phase optimization, binary amplitude optimization and multi-objective optimization are also numerically proved to be benefit from GM. In general, with the introduction of GM, one can easily and effectively promote personal optimization system built by iterative optimization algorithm. It can be anticipated that, not only for iterative WFS, GM is potentially contributed to a variety of studies that aim to find the optimal solution in complex model, such as manipulating multi-dimensional characteristics of the fiber laser [17, 18], two-photon microscopy [19].

## 2. Principle and simulation

In WFS, the coherent light is modulated using a spatial light modulator (SLM) or digital micromirror device (DMD) before being transmitted through scattering medium. To focus through scattering medium, iterative algorithms are utilized to search for the optimal phase or amplitude mask, which contribute to the optimum enhancement of the optical focus. An iterative algorithm typically operates by initialization, evaluation, evolution and selection, as shown in Fig.1(a). In a GM-enhanced iterative algorithm, GM will function once the regular evolution process has finished, as denoted by the dashed frame. The diagram of GM is illustrated in Fig.1 (b) and the detailed steps are as follows:

Step 1: Calculate guidance map from the current population consists of $NP$ individuals. The guidance map $G$ is defined as a $N \times L$ matrix for $N$ input modes (pixels) and $L$ available pixel values. In the $k^{th}$ generation, the guidance $G_{n,l}$ for the $n^{th}$ pixel taking the $l^{th}$ pixel value can be calculated according to

$$\begin{cases} G_{n,l}^k = \sum_{i=1}^{NP} \frac{f_i \cdot \delta_{n,l}}{Rank_i} \\ \delta_{n,l} = \begin{cases} 1 & pop_{i,n}^k == D_l \\ 0 & Otherwise \end{cases} \end{cases}, \quad (1)$$

where $f_i$ and $Rank_i$ are the fitness value and rank index of the $i^{th}$ individual. A lower rank index corresponds to an individual with higher fitness. $pop_{i,n}^k$ is the pixel value of the $n^{th}$ mode of the $i^{th}$ individual in the $k^{th}$ generation. We note that in the phase modulation, the pixel value is $D_l \in (0, \frac{2\pi}{L}..., 2\pi)$, while $D_l \in (0,1)$ is available in the binary amplitude modulation.

Step 2: Randomly select number of $N_G$ guided pixels from the $i^{th}$ individual $pop_i^k$ in the $k^{th}$ generation, following

$$N_G = N \cdot \left[ (G_0 - G_{end})^{\frac{k}{\lambda_G}} + G_{end} \right],$$ where $G_0$ and $G_{end}$ are the initial and final GM rate, $\lambda_G$ is the decay factor of GM rate.

Step 3: Impose guidance and obtain $NS$ guided updated individuals, in which $N_G$ pixels of each updated individual are guided by $G_{N \times L}$.

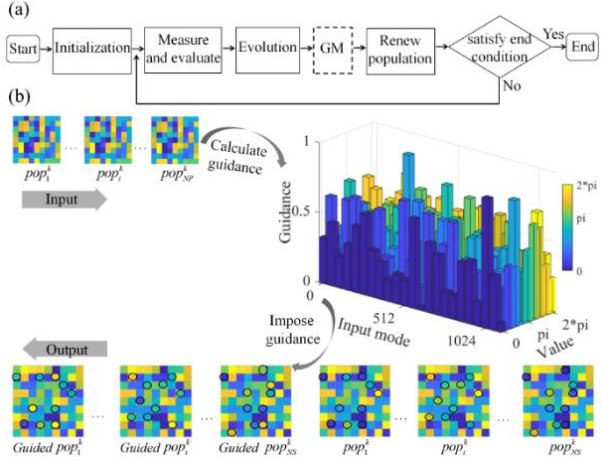

Fig. 1. (a) Universal flow chart of GM-enhanced WFS optimization algorithm. (b) The diagram of GM.

Table 1 Parameters for different algorithms[a]

| Algorithms | GA | PSO | ACO | SA |
|---|---|---|---|---|
| Regular methods | $N = 1024$<br>$NP = 36$<br>$NS = 18$<br>$L = 20$<br>$R_0 = 0.1$<br>$R_{end} = 0.0025$<br>$\lambda_m = 250$ | $N = 1024$<br>$NP = 36$<br>$NS = 18$<br>$L = 20$<br>$c_1 = 4$<br>$c_2 = 4$<br>$w = 0.9$<br>$w_d = 0.99$ | $N = 1024$<br>$NP = 36$<br>$NS = 18$<br>$L = 20$<br>$\alpha = 0.9$ | $N = 1024$<br>$L = 20$<br>$Pn = N/2$<br>$Pv = \pi/20$<br>$T_0 = 1$<br>$T_d = 0.99$<br>$Pl = N/16$ |
| GM-enhanced | $G_0 = 0.1$; $G_{end} = 0.0025$; $\lambda_G = 250$; | | | |

[a] $N$, input mode number; $NP$, population size; $NS$, the number of updated individuals in each generation; $L$, the number of pixel value; $R_0$, initial mutation rate, $R_{end}$, final mutation rate; $\lambda_m$, mutation decay constant; $c_1$, individual learning factor; $c_2$, social learning factor; $w$, weight factor; $w_d$, weight decay constant; $\alpha$, pheromone decay constant; $Pn$, perturbated pixels number; $Pv$, perturbation value; $T_0$, initial temperature; $T_d$, temperature decay constant; $Pl$, perturbation loop length.

To validate the benefits of GM, single point focusing through scattering medium is numerically simulated by GA, PSO, ACO, SA and the corresponding GM-enhanced algorithms. The detailed procedures of these algorithms can be found in Supplement 1. The well-tuned parameters used for these algorithms are listed in Table 1. The enhancement factor, defined as the ratio between the intensity of focus after optimization and the initial averaged intensity of the speckle pattern before optimization, is the common fitness function. Figure 2 presents the enhancement factor as a function of the number of measurements, after taking the average of 10 repeated calculations of each algorithm. The optimization efficiency herein is evaluated by the enhancement factor after optimization and enhancing rate. It can be seen that all the GM-enhanced methods obtain larger enhancement factors than the corresponding regular method without GM. Specifically, with the assistance of GM, GA-GM, PSO-GM, ACO-GM, SA-GM reach final enhancement factor at 871, 772, 801, 742, indicating an improvement of 13%, 25%, 240%, 17% over the regular ones respectively. Moreover, the measurement at which GM-enhanced method reaches the maximum enhancement factor of regular one is picked to metric the enhancing rate. As denoted by the marked coordinate in each optimization curve, the maximum enhancement factor achieved by regular methods require only 28260, 21456, 1908 and 12528 measurements for the GM-enhanced methods. As a result of GM, GM-enhanced methods release 69%, 76%, 98% and 86% redundant measurements, which significantly boost the enhancing rate of the regular methods.

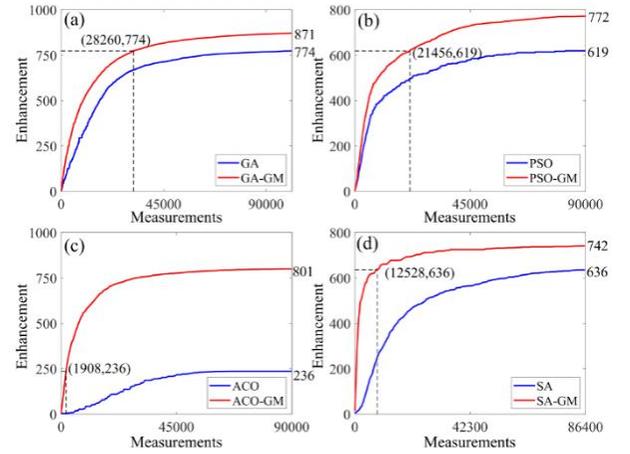

Fig. 2. Simulated optimization process of (a) GA and GA-GM; (b) PSO and PSO-GM; (c) ACO and ACO-GM; (d) SA and SA-GM.

3. Experimental results and analysis

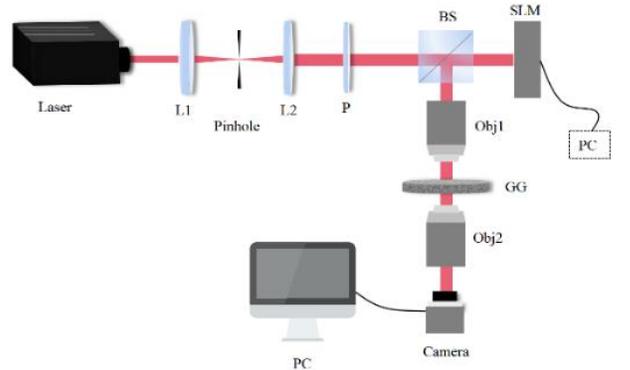

Fig. 3. The experimental setup for iterative WFS.

Iterative WFS experiments are conducted with the setup shown in Fig. 3. The laser beam (1064 nm) is filtered and collimated by a telescope formed by lens L1, pinhole and lens L2. This plane wave incidents on the SLM (Hamamatsu X13138-03WR), which is a phase only SLM of 1272×1024 pixels with a pitch size of 12.5 µm. During the experiment, 32×32 input modes are used, with 16×16 pixels grouped as a mode. A linear polarizer (P) is placed in front of the SLM to ensure the polarization state of the input light to match the polarization sensitivity of the SLM. The modulated light is reflected by the beam splitter (BS) and focused by the objective (Obj1, 40 NA=0.65) into the ground glass (GG, Thorlabs DG10-220, 220 grit, 2 mm thickness). Another objective (Obj2, 25 NA=0.4) is placed behind GG to collect scattered light. A camera (Thorlabs, CS2100M) records the intensity of the scattered field. The control of the SLM and camera is performed by the computer (PC).

With the experimental setup, the regular algorithms and the GM-enhanced ones are both validated experimentally for single-point focusing through scattering medium with the parameters as listed in Table 1. Figure 4 depicts the evolution of enhancement factor as a function of the measurement. It is evident from the experimental results that, in agreement with the simulation results, all the GM-enhanced methods gain a greater enhancement factor than their corresponding regular ones. The enhancement factors achieved by GA-GM, PSO-GM, ACO-GM and SA-GM are, respectively, 388, 345, 326 and 368, which are, respectively, 11%, 28%, 189% and 16% higher than the corresponding regular one. Additionally, we also mark the measurement at which the GM-enhanced method achieves the maximum enhancement as made by regular one finally. As denoted by these marked coordinates, GA-GM, PSO-GM, ACO-GM, SA-GM reach the maximum enhancement factor of corresponding regular method at the $28998^{th}$, $19224^{th}$, $1422^{th}$, $17604^{th}$ measurement respectively, which indicates that GM-enhanced methods release 46%, 64%, 97%, 68% redundant measurements. Moreover, Figure 5 presents the normalized speckle intensity images after optimization using the four regular algorithms and GM-enhanced ones, the red lines plot the horizontal profiles of the focus spot. As can be observed, the final intensity distribution optimized with GM-enhanced methods exhibit brighter focusing spot and greater background suppression due to the introduction of GM.

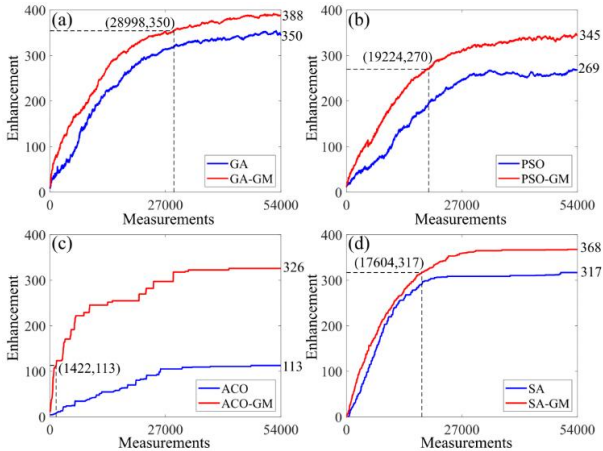

Fig. 4. Experimental optimization process of (a) GA and GA-GM; (b) PSO and PSO-GM; (c) ACO and ACO-GM; (d) SA and SA-GM.

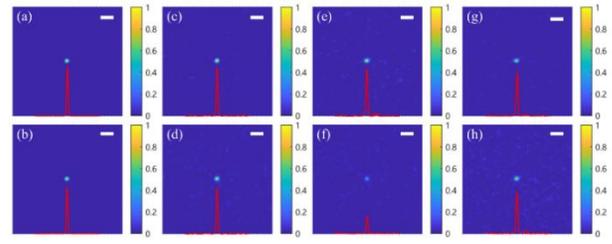

Fig. 5. The normalized speckle intensity images after optimization by (a) GA-GM; (b) GA; (c) PSO-GM; (d) PSO; (e) ACO-GM; (f) ACO; (g) SA-GM; (h) SA in experiment. The scale bar represents 150 µm.

In addition to the above 32×32 input mode setting, 16×16 and 64×64 setting with 32×32 and 8×8 pixels grouped as a macro-pixel are also implemented in simulation and experiment with the same parameters in Table 1. Table 2 lists the final improvements of the four GM-enhanced methods over the regular ones for different input mode numbers. It can be observed that the improvements in 16×16 input modes are rather modest, however, With the increase of the number of input modes, the improvements become more significant. The improvements in 64×64 input modes are no less than 20% in both simulated and experimental tests. The comparison between different input mode numbers suggests that GM will bring more significant improvements with a larger number of input mode.

Table 2 The improvements of four algorithms for different input mode numbers

|  | Simulation results | | | Experimental results | | |
| --- | --- | --- | --- | --- | --- | --- |
|  | 16×16 | 32×32 | 64×64 | 16×16 | 32×32 | 64×64 |
| GA-GM | 5% | 13% | 21% | 4% | 11% | 20% |
| PSO-GM | 9% | 25% | 77% | 12% | 28% | 79% |
| ACO-GM | 89% | 240% | 550% | 56% | 189% | 511% |
| SA-GM | 8% | 17% | 29% | 7% | 16% | 30% |

To further explore the universal benefits of GM, binary amplitude optimization and multi-objective optimization (MOO) are also investigated. As is well known, leveraging the high frame rate of DMD, binary amplitude modulation is widely employed in WFS. As a typical instance, binary GA and GA-GM are numerically simulated with the same parameters in Table 1, except $L = 2$. The ratio of the focal spot and average background intensity of the speckle (PBR) is set as the shared fitness function of binary GA and GA-GM. As shown in Fig. 6 (a), binary GA finally traps into a local optimum at 175, however GA-GM reach optimized PBR at 185, indicating an improvement of 6%. Moreover, the marked coordinate suggests binary GA-GM reach the maximum PBR achieved by GA at $12528^{th}$ measurement, which reduce 77% redundant measurements.

In some specific research fields, MOO is more effective than single objective optimization [20, 21]. A multi-point focusing through scattering medium with the classic multi-objective genetic algorithm (NSGAII) [22] is numerically analyzed with same parameters of GA and GA-GM in Table 1. Its procedure can be found in supplement 1. In this MOO problem, average enhancement and uniformity are two common fitness functions. Average enhancement is defined as the ratio of the average intensity of the multiple focal spots to the average intensity before optimization, and uniformity is defined as the standard deviation of the multiple focal spots [23]. Fig. 6 (b) presents the average enhancement and uniformity of six focus spots as the function of measurements, the zoomed-in inset exhibits the converged results of uniformity in black frame. It can be seen that NSGAII-GM achieves a higher and more uniform enhancement than NSGAII, with an improvement of 13% on average enhancement and 200% on uniformity. Similar to the single objective optimization, GM offers an enhanced effect for NSGAII in MOO.

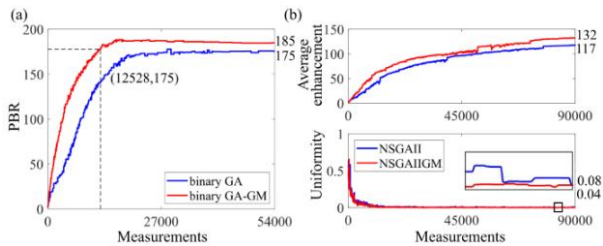

Fig. 6. Simulated optimization process of (a) binary GA and binary GA-GM; (b) NSGAII and NSGAII-GM.

From the above results and analysis, GM effectively improves the optimization efficiency in terms of optimization results and rate for the commonly-used iterative algorithms. With the increase of the number of input modes, the improvement effect of GM becomes more significant. Moreover, the investigation of binary and multi-objective optimization further reveals that GM can be applied to binary amplitude optimization system and multi-objective optimization. In a word, these results and analysis in simulation and experiments demonstrate the improvements and university of GM.

## 4. conclusion

In conclusion, we propose a mutation strategy with guidance for iterative WFS algorithms. The proposed method effectively enhances the optimization performance for various algorithms. More than the improvements generalized in various iterative algorithms, different modulation methods including phase or binary amplitude modulation demonstrate the advantages of GM. It can be reasonable to anticipate that GM is promising to bring essential assistance for broad community of WFS and wider applications, ranging from controlling the transmission of light through disorder media to optical-manipulation behind them.

### Acknowledgement

This work was supported by the National Natural Science Foundation of China (NSFC) (No. 61605049), The Youth Innovation Foundation of Xiamen City (No. 3502Z20206013); The Fundamental Research Funds for the Central Universities (No. ZQN707).

### Supplement document

See Supplement 1 for more detailed flowcharts of algorithms.

### References

1. I. M. Vellekoop, "Feedback-based wavefront shaping," Opt Express **23,** 12189 (2015).
2. D. Wu, J. Luo, Z. Li, and Y. Shen, "A thorough study on genetic algorithms in feedback-based wavefront shaping," Journal of Innovative Optical Health Sciences **12,** 1942004 (2019).
3. X. Zhang and P. Kner, "Binary wavefront optimization using a genetic algorithm," Journal of Optics **16,** 125704 (2014).
4. C. Z.-Y. HUANG Hui-Ling, SUN Cun-Zhi, LIU Ji-Lin, PU Ji-Xiong, "Light Focusing through Scattering Media by Particle Swarm Optimization," Chin. Phys. Lett., 104202%V 32 (2015).
5. L. Fang, H. Zuo, Z. Yang, X. Zhang, L. Pang, W. Li, Y. He, X. Yang, and Y. Wang, "Particle swarm optimization to focus coherent light through disordered media," Applied Physics B **124,** 155 (2018).
6. Z. Fayyaz, N. Mohammadian, F. Salimi, A. Fatima, M. R. R. Tabar, and M. R. N. Avanaki, "Simulated annealing optimization in wavefront shaping controlled transmission," Applied Optics **57,** 6233 (2018).
7. F. Zahra, S. Faraneh, M. Nafiseh, F. Afreen, M. R. R. Tabar, and R. N. A. Mohammad, "Wavefront shaping using simulated annealing algorithm for focusing light through turbid media," in *Proc.SPIE*, (2018), 104946M.
8. Z. Yang, L. Fang, X. Zhang, and H. Zuo, "Controlling a scattered field output of light passing through turbid medium using an improved ant colony optimization algorithm," Optics and Lasers in Engineering **144,** 106646 (2021).
9. L. Fang, H. Zuo, Y. Xu, and B. Ma, "Focusing light through scattering media by combining genetic and Gauss–Newton algorithms," Applied Physics B **125,** 94 (2019).
10. Y. Luo, S. Yan, H. Li, P. Lai, and Y. Zheng, "Focusing light through scattering media by reinforced hybrid algorithms," APL Photonics **5,** 016109 (2020).
11. B.-Q. Li, B. Zhang, Q. Feng, X.-M. Cheng, Y.-C. Ding, and Q. Liu, "Shaping the Wavefront of Incident Light with a Strong Robustness Particle Swarm Optimization Algorithm," Chinese Physics Letters **35,** 124201 (2018).
12. Q. Feng, B. Zhang, Z. Liu, C. Lin, and Y. Ding, "Research on intelligent algorithms for amplitude optimization of wavefront shaping," Applied Optics **56,** 3240 (2017).
13. C. M. Woo, Q. Zhao, T. Zhong, H. Li, Z. Yu, and P. Lai, "Optimal efficiency of focusing diffused light through scattering media with iterative wavefront shaping," APL Photonics **7,** 046109 (2022).
14. C. M. Woo, H. Li, Q. Zhao, and P. Lai, "Dynamic mutation enhanced particle swarm optimization for optical wavefront shaping," Opt Express **29,** 18420 (2021).
15. H. Li, C. M. Woo, T. Zhong, Z. Yu, Y. Luo, Y. Zheng, X. Yang, H. Hui, and P. Lai, "Adaptive optical focusing through perturbed scattering media with a dynamic mutation algorithm," Photonics Research **9,** 202 (2021).
16. Q. Zhao, C. M. Woo, H. Li, T. Zhong, Z. Yu, and P. Lai, "Parameter-free optimization algorithm for iterative wavefront shaping," Optics Letters **46,** 2880 (2021).
17. X. Wei, J. C. Jing, Y. Shen, and L. V. Wang, "Harnessing a multi-dimensional fibre laser using genetic wavefront shaping," Light: Science & Applications **9,** 149 (2020).
18. O. Tzang, A. M. Caravaca-Aguirre, K. Wagner, and R. Piestun, "Adaptive wavefront shaping for controlling nonlinear multimode interactions in optical fibres," Nature Photonics **12,** 368 (2018).
19. J. Tang, R. N. Germain, and M. Cui, "Superpenetration optical microscopy by iterative multiphoton adaptive

compensation technique," Proceedings of the National Academy of Sciences **109,** 8434 (2012).
20. H. Li, X. Wu, G. Liu, R. V. Vinu, X. Wang, Z. Chen, and J. Pu, "Generation of controllable spectrum in multiple positions from speckle patterns," Optics & Laser Technology **149,** 107820 (2022).
21. S. Honda, T. Igarashi, and Y. Narita, "Multi-objective optimization of curvilinear fiber shapes for laminated composite plates by using NSGA-II," Composites Part B: Engineering **45,** 1071 (2013).
22. K. Deb, A. Pratap, S. Agarwal, and T. Meyarivan, "A fast and elitist multiobjective genetic algorithm: NSGA-II," IEEE. Trans. Evol. Comput. 6 (2), 182-197 (2002).
23. Q. Feng, F. Yang, X. Xu, B. Zhang, Y. Ding, and Q. Liu, "Multi-objective optimization genetic algorithm for multi-point light focusing in wavefront shaping," Opt Express **27,** 36459 (2019).